\documentstyle[aps,12pt]{revtex}
\newcommand{\bb}{\begin{equation}}
\newcommand{\ee}{\end{equation}}
\newcommand{\ba}{\begin{array}}
\newcommand{\ea}{\end{array}}

\begin{document}
\baselineskip 2.25pc

\title{Nonlocal scalar electrodynamics from Chern-Simons theory}
\author{Qiong-gui Lin\thanks{Electronic address: stdp@zsu.edu.cn}}
\address{Department of Physics, Zhongshan University, Guangzhou
        510275,\\
        People's  Republic of China\thanks{Mailing address}\\
        and\\
        China Center of Advanced Science and Technology (World
	Laboratory),\\
        P.O.Box 8730, Beijing 100080, People's Republic of China}

\maketitle
\vfill

\begin{abstract}
{\normalsize
	The theory of a complex scalar interacting with a pure Chern-Simons
gauge field is quantized canonically. Dynamical and nondynamical variables
are separated in a gauge-independent way. In the physical subspace of the
full Hilbert space, this theory reduces to a pure scalar theory with nonlocal
interaction. Several scattering processes are studied and the cross sections
are calculated.}
\end{abstract}
\vfill

\leftline {PACS number(s): 11.15.-q, 11.10.Ef, 11.10.Lm}
\newpage

	Chern-Simons theories were widely studied in recent years. It has
become familiar to both field theorists and condensed matter physicists
in these days. To field theorists, of particular interest is the
Chern-Simons theory without a Maxwell term [1].
In this field of investigation,
much attention has been focused on classical soliton solutions [2-6] and
the problem of fractional spin and exotic statistics [7,8]. To our knowledge
the quantum processes of this theory has not been studied in the
manner presented here. We quantize the theory by Dirac's method [9]. By
using some appropriate canonical transformations, dynamical and
nondynamical degrees of freedom are separated in a gauge-independent
way [10]. After quantization, this theory naturally reduces to a pure
scalar theory with nonlocal interaction in the physical subspace of the
full Hilbert space. On the basis of this argument, several two-body
scattering processes are studied and the corresponding
cross sections are obtained.

	Let us begin with the Lagrangian density
$${\cal L}=(D_\mu\varphi)^*D^\mu\varphi -m^2\varphi^*\varphi +
{\kappa\over 2}\epsilon^{\lambda\mu\nu}A_\lambda\partial_\mu A_\nu \eqno(1)$$
where $D_\mu=\partial_\mu+ieA_\mu$. It describes a massive complex scalar
field $\varphi $ with mass $m$ interacting with a Chern-Simons gauge field
$A_\mu$. The canonical momenta are given by
$$\pi={\partial{\cal L}\over \partial \mathaccent 95  \varphi }
=(D_0\varphi)^*,
\quad\pi^*={\partial{\cal L}\over \partial \mathaccent 95  \varphi^*}
=D_0\varphi,
\eqno(2a)$$
$$\pi_0={\partial{\cal L}\over \partial \mathaccent 95  A^0 }=0,
\quad\pi_i={\partial{\cal L}\over \partial \mathaccent 95  A^i }=
-{\kappa\over 2}\epsilon_{ij}A_j.\eqno(2b)$$
The nonvanishing Poisson brackets among the canonical variables read
$$\{\varphi({\bf x},t),\pi({\bf y},t)\} =\delta({\bf x}-{\bf y}),
\quad\{\varphi^*({\bf x},t),\pi^*({\bf y},t)\} =\delta({\bf x}-{\bf y}),
\eqno(3a)$$
$$\{A_\mu({\bf x},t),\pi_\nu({\bf y},t)\} =g_{\mu\nu}\delta({\bf x}-{\bf y}).
\eqno(3b)$$
Eqs.(2b) give the following primary constraints.
$$\pi_0(x)\approx0,\eqno(4a)$$
$$\chi_i(x)\equiv\pi_i(x)+{\kappa\over 2}\epsilon_{ij}A_j(x)\approx 0.
\eqno(4b)$$
After dropping some surface terms and taking Eqs.(4) into account, the
canonical Hamiltonian takes the form
$$ H_c=\int d{\bf x}\;[\pi^*\pi+ (D_i\varphi)^*D_i\varphi +m^2\varphi^*
\varphi+ A_0(\rho+\kappa B)
+\lambda_0\pi_0+\lambda_i\chi_i]\eqno(5)$$
where $\rho=ie(\pi^*\varphi^*-\pi\varphi )$ and $ B=-\epsilon_{ij}\partial_i
 A_j$ are the charge density and magnetic field, respectively, and
$\lambda_0$ and $\lambda_i$ are Lagrange multipliers
which may be functions of
the canonical variables. The consistency conditions
$\mathaccent 95  \pi_0\approx0$ and $\mathaccent 95  \chi_i\approx0$
lead to
$$\xi(x)\equiv\rho(x)+\kappa B(x)\approx0,\eqno(6)$$
$$\lambda_i(x)=-\partial_i A_0(x)+{1\over\kappa}\epsilon_{ik}J_k(x).
\eqno(7)$$
Eq.(6) is a secondary constraint while Eq.(7) gives $\lambda_i$ in terms
of the canonical variables.
In Eq.(7) $J_k$ are the  spatial components of the conserved current
$$ J_\mu=ie[\varphi^*D_\mu \varphi-\varphi (D_\mu\varphi)^*]
\eqno(8)$$
whose time component $J_0=\rho$ has been mentioned above. The further
consistency condition $ \mathaccent 95	\xi\approx0$ yields
$$ \kappa\epsilon_{ij}\partial_i\lambda_j+\partial_i J_i=0\eqno(9) $$
which, on account of (7), is satisfied. Thus Dirac's algorithm ends
and all the constraints are given by (4) and (6). Eq.(4a) is obviously
a first-class constraint, while (4b) and (7), though at first sight
are all of second class, can be linearly combined to give a first-%
class one [11]:
$$\mathaccent "7E\xi(x)=\xi(x)+\partial_i\chi_i(x).\eqno(10)$$
Eq.(4b) is really of second class. It is easy to show that
$$C_{ij}({\bf x},{\bf y})\equiv\{\chi_i({\bf x} ,t),\chi_j({\bf y},t)\}
=-\kappa\epsilon_{ij}\delta({\bf x} -{\bf y}), \eqno(11a)$$
$$C^{-1}_{ij}({\bf x},{\bf y} )={1\over\kappa}\epsilon_{ij}\delta
({\bf x} -{\bf y}) . \eqno(11b)$$
Following the standard definition the nonvanishing Dirac brackets can
be shown te be
$$\{\varphi({\bf x},t),\pi({\bf y},t)\}^* =\delta({\bf x}-{\bf y}),
\quad\{\varphi^*({\bf x},t),\pi^*({\bf y},t)\}^* =\delta({\bf x}-{\bf y}),
\eqno(12a)$$
$$\{A_0({\bf x},t),\pi_0({\bf y},t)\}^* =\delta({\bf x}-{\bf y}),
\eqno(12b)$$
which are simply the same as the Poisson ones and
$$\{A_i( {\bf x},t),A_j({\bf y},t)\}^*={1\over\kappa}\epsilon_{ij}
\delta({\bf x} -{\bf y}) .\eqno(12c)$$
Also nonvanishing are $ \{A_i( {\bf x},t),\pi_j({\bf y},t)\}^*$
and $\{\pi_i( {\bf x},t),\pi_j({\bf y},t)\}^* $ which are not useful
henceforce and are thus not given here. After the Poisson brackets
are replaced by the Dirac ones, Eq.(4b)  becomes a strong equation and
thus $\mathaccent "7E\xi(x)=\xi(x)$. The total Hamiltonian involves
all the first-class constraints (primary and secondary) and takes
the form
$$ H_T=\int d{\bf x}\;[\pi^*\pi+ (D_i\varphi)^*D_i\varphi +m^2\varphi^*
\varphi
+\lambda_0\pi_0+\lambda\xi],\eqno(13)$$
where we have included the $A_0$ appearing in Eq.(5) in the Lagrange
multiplier $\lambda$. The equation of motion for any canonical variable
$g$ reads
$$\mathaccent 95 g=\{g,H_T\}^*.\eqno(14)$$
This can also be obtained via variation from the first-order Lagrangian
$$L=\int d{\bf x}\;\left[\pi\mathaccent 95\varphi+\pi^*\mathaccent 95
\varphi^*+{\kappa\over 2}\epsilon_{ij}A_j\mathaccent 95  A_i
+\pi_0\mathaccent 95  A_0\right]-H_T.\eqno(15)$$
Variations of $L$ with respect to $\lambda_0$ and $\lambda$ lead to the
two first-class constraints. From (15) we see that the canonical pairs
are $(\varphi,\pi)$, $(\varphi^*,\pi^*)$,$(A_0,\pi_0)$, and $(A_1,\kappa
A_2)$. The last one is not convenient. Thus we decompose $A_i$ into its
longitudinal and transverse components as
$$A_i=A_i^L+A_i^T\equiv\partial_i\omega+\epsilon_{ij}\partial_j\eta.
\eqno(16)$$
It is easy to show that $\nabla^2\eta=B$, and the only nonvanishing
Dirac brackets among $\omega(x)$ and $B(x)$ is
$$\{\omega({\bf x},t),\kappa B({\bf y},t)\}^*=\delta({\bf x} -{\bf y} ).
\eqno(17)$$
Substituting (16) into (15) and dropping some surface terms we have
$$L=\int d{\bf x}\;[\pi\mathaccent 95\varphi+\pi^*\mathaccent 95
\varphi^*+\kappa B\mathaccent 95 \omega
+\pi_0\mathaccent 95  A_0]-H_T\eqno(18)$$
where $A_i$ in $H_T$ is expressed in $\omega$ and $B$ by using the relation
$$A_i^T({\bf x} ,t)=\int d{\bf y}\;\epsilon_{ij}\partial_jG({\bf x} -{\bf y})
B({\bf y},t)\eqno(19)$$
where the partial derivative $\partial_j$ is with respect to {\bf x}
and the Green function $G({\bf x}) ={1\over 2\pi}\ln|{\bf x} |$
satisfies $\nabla^2G({\bf x} )=\delta({\bf x} )$. Now the canonical pair
$(A_1,\kappa A_2)$ is replaced by $(\omega,\kappa B)$. We further make the
canonical transformation [10]
$$\varphi=\exp(-ie\omega)\phi,\quad \pi=\exp(ie\omega)\Pi,\eqno(20a)$$
$$\varphi^*=\exp(ie\omega)\phi^*,\quad \pi^*
=\exp(-ie\omega)\Pi^*,\eqno(20b)$$
then the Lagrangian becomes
$$L=\int d{\bf x}\;[\Pi\mathaccent 95\phi+\Pi^*\mathaccent 95
\phi^*+\xi\mathaccent 95 \omega
+\pi_0\mathaccent 95  A_0]-H_T\eqno(21)$$
where
$$ H_T=\int d{\bf x}\;[\Pi^*\Pi+ (D_i^T\phi)^*D_i^T\phi +m^2\phi^*
\phi
+\lambda_0\pi_0+\lambda\xi]\eqno(22a)$$
where $D_i^T=\partial_i+ieA_i^T$ and
$$ A_i^T({\bf x},t) ={1\over \kappa}\int d{\bf y} \;\epsilon_{ij}
\partial_j G({\bf x} -{\bf y})[\xi({\bf y},t)-\rho({\bf y},t)]
\eqno(23)$$
where $\rho=ie[\Pi^*\phi^*-\Pi\phi]$. It is easy to show that the only
nonvanishing Dirac brackets among the new canonical variables are
$$\{\phi({\bf x},t),\Pi({\bf y},t)\}^* =\delta({\bf x}-{\bf y}),\quad
\{\phi^*({\bf x},t),\Pi^*({\bf y},t)\}^* =\delta({\bf x}-{\bf y}),
\eqno(24a)$$
$$\{A_0({\bf x},t),\pi_0({\bf y},t)\}^* =\delta({\bf x}-{\bf y}),
\eqno(24b)$$
$$\{\omega( {\bf x},t),\xi({\bf y},t)\}^*=
\delta({\bf x} -{\bf y}) .\eqno(24c)$$
Therefore the new canonical pairs are
$(\phi,\Pi)$, $(\phi^*,\Pi^*)$,$(A_0,\pi_0)$, and $(\omega,\xi)$.
It is remarkable that the first-class constraint $\xi$ becomes a canonical
momentum. Denote $\Phi=(\phi,\Pi,\phi^*,\Pi^*)$. Since $\Phi$ have vanishing
Dirac brackets with $\xi$, they are gauge invariant (under time-independent
gauge transformations). One easily realizes that $A_0$ and $\omega$
are not dynamical variables since their conjugate momentums vanish.
By redefining $\lambda$ in the Hamiltonian (22a) Eq.(23) can be replaced by
$$ A_i^T({\bf x},t) =-{1\over \kappa}\int d{\bf y} \;\epsilon_{ij}
\partial_j G({\bf x} -{\bf y})\rho({\bf y},t).
\eqno(22b)$$

	We are now in a position to quantize the theory described by
(22) and (24). This is accomplished by promoting the Dirac brackets
to commutators. Thus (24a), say, becomes
$$[\phi({\bf x},t),\Pi({\bf y},t)] =i\delta({\bf x}-{\bf y}),
[\phi^\dagger({\bf x},t),\Pi^\dagger({\bf y},t)]=i\delta({\bf x}-{\bf y})
\eqno(25)$$
after quantization. Here $\phi^\dagger$ is the Hermitian conjugate of
$\phi$, replacing the complex conjugate $\phi^*$ in the classical theory.
Meanwhile the equation of motion for any canonical variable $g$ is
changed to
$$ i\mathaccent 95 g=[g,H_T],\eqno(26)$$
where $H_T$ involves ordering ambiguity in the quantized theory and should
be handled carefully. To complete the story of quantization, the two
first-class constraints are realized as weak equations at the quantum
level:
$$\pi_0(x)|{\rm phys}\rangle=0, \xi(x)|{\rm phys}\rangle =0\eqno(27)$$
where $|{\rm phys}\rangle$ represents a physical state.

	We emphasize that in the quantum theory it is sufficient to deal
with the system described by (22).
There is no need to go back to the original
one described by (13). It also seems that $\phi$ etc. are more ``physical''
than $\varphi$ etc. since the former are gauge invariant.

	Any physical quantity $F$ should be independent of $A_0$ and $\omega$
since they are pure gauge variables. If $F$ depends on $\pi_0$ and $\xi$
(such as the Hamiltonian), it has in general the form $F(\Phi,\pi_0,\xi)$.
Since $\pi_0$ and $\xi$ commute with $\Phi$, and a physical state satisfies
(27),  one always has
$$ F(\Phi,\pi_0,\xi)|{\rm phys}\rangle =F(\Phi,0,0)|{\rm phys}\rangle.
\eqno(28)$$
Thus in the physical subspace of the full Hilbert space, $\pi_0$ and
$\xi$, when involved in any physical quantity, can be identically set to zero.
Therefore only $\Phi$ are physically relevant.
As the time evolution of $\Phi$
is independent of the last two terms in (22a),
these two terms can be dropped.
Then the theory described by (22) and (25) reduces in the physical subspace
to a pure scalar theory
with nonlocal interaction.

	Physical states of the theory can be constructed from canonical
variables and the vacuum state $|0\rangle$ satisfying (27):
$$\pi_0(x)|0\rangle=0, \xi(x)|0\rangle =0.\eqno(29)$$
This implies that the vacuum state is independent of $A_0$ and $\omega$,
which seems quite natural. Since $\Phi$ commute with $\pi_0$ and $\xi$,
any state generated by operating $\Phi$ or functions of them on
$|0\rangle$ automatically satisfies (27) and is thus physical.
On the other hand, operating $\omega$ or $A_0$ on $|0\rangle$ can not
generate a physical state since it fails to satisfy (27). In other
words, physical states can only be generated by $\Phi$ and their functions.
This further confirms the statement that only $\Phi$ are physically relevant.

	In the following we decompose $H_T$ into two parts:
$$ H_T=H_0+H\eqno(30a)$$
where
$$ H_0=\int d{\bf x}\;:[\Pi^\dagger\Pi+\partial_i\phi^\dagger\partial_i\phi
+ m^2\phi^\dagger\phi]:,\eqno(30b)$$
$$ H= \int d{\bf x}\;:[ie(\partial_i\phi^\dagger A_i^T\phi-\phi^\dagger
A_i^T\partial_i\phi)+e^2\phi^\dagger A_i^T A_i^T\phi]:.\eqno(30c)$$
$H_0$ and $H$ are respectively the free Hamiltonian and the interacting
Hamiltonian. $H$ involves ordering ambiguity and we have adopted the
normal-ordering prescription denoted by colons,
which will become clear in the
interaction picture (see below).
Normal ordering of $H_0$ is not necessitated
by ordering ambiguity. It is just employed to discard the zero-point energy.
In Eq.(30) $A_i^T$ is given by (22b) with
$$\rho=ie:[\Pi^\dagger\phi^\dagger-\Pi\phi]:\eqno(31)$$
such that $\rho$ as well as $A_i^T$ is Hermitian. Normal ordering of $\rho$
also act to remove a zero-point charge.

	 After making the decomposition (30),
we go to the interaction picture (see for example Ref.[12]) where $\phi$,
$H$ etc. are transformed to $\phi_I$, $H_I$ etc..
In the following we work in
the interaction picture but omit the subscript $I$. In this picture the
commutation relation (25) remains unchanged while the time evolution of a
canonical variable $g$ and a physical state $|\psi\rangle$
are governed by
$$i\mathaccent 95 g=[g,H_0],\eqno(32)$$
$$i{\partial\over\partial t}|\psi\rangle=H|\psi\rangle.\eqno(33)$$
From (32) and (30b) we have
$$ \Pi=\mathaccent 95 \phi^\dagger,\quad \Pi^\dagger=\mathaccent 95 \phi,
\eqno(34a)$$
$$ (\partial_\mu^2+m^2)\phi=0,\quad  (\partial_\mu^2+m^2)\phi^\dagger=0.
\eqno(34b)$$
Thus both $\phi$ and $\phi^\dagger$ obey the Klein-Gordon equation, and
we have the following expansions.
$$\phi(x)=\sum_{\bf k}{1 \over\sqrt{2 k_0 V}}[a_{\bf k}e^{-ik\cdot x}
+b_{\bf k}^\dagger e^{ik\cdot x}],\eqno(35a)$$
$$\Pi(x)=\sum_{\bf k}i\sqrt{k_0\over 2V} [a_{\bf k}^\dagger e^{ik\cdot x}
-b_{\bf k} e^{-ik\cdot x}]\eqno(35b)$$
where $V$ is a two-dimensional normalization volume and similar ones
for $\phi^\dagger$ and $\Pi^\dagger$. In (35)
$$ k_0=\sqrt{{\bf k}^2+m^2}.\eqno(36)$$
Using Eq.(25), one can show that the nonvanishing commutators among
$a_{\bf k} $, $b_{\bf k} $ etc. are given by
$$ [a_{\bf k} , a_{\bf l}^\dagger]=\delta_{\bf kl},\quad
[b_{\bf k} , b_{\bf l}^\dagger]=\delta_{\bf kl}.\eqno(37)$$
The vacuum state is defined by
$$a_{\bf k}|0\rangle=b_{\bf k}|0\rangle=0, \quad\forall{\bf k} \eqno(38)$$
in addition to be independent of $A_0$ and $\omega$.
Then it can be shown that both $a_{\bf k}^\dagger|0\rangle$ and
$b_{\bf k}^\dagger|0\rangle$ are one-particle states carrying momentum
{\bf k} and energy $k_0$. The former carries charge $e$ while the latter
carries $-e$. Thus $a_{\bf k}^\dagger$ may be regarded as the creation
operator of a particle while $b_{\bf k}^\dagger$ of an antiparticle.
From (33), it can be shown that the amplitude of transition for the
system from an initial state $|i\rangle$ at $t=-\infty$ to a final one
$|f\rangle$ at $t=+\infty$ is
$$A_{fi}=\langle f|S|i\rangle \eqno(39)$$
where $S$ is the $S$-matrix given by
$$ S=\sum_{n=0}^\infty S^{(n)},\eqno(40a)$$
$$ S^{(n)}={(-i)^n\over n!}\int_{-\infty}^{+\infty} dt' dt''
\cdots dt^{(n)}\;
T[H(t')H(t'')\cdots H(t^{(n)})].\eqno(40b)$$
The lowest-order contribution to scattering processes comes from $S^{(1)}$,
and for two-body scattering the last term in (30c) does not contribute
(to this order).
Thus for two-body scattering we have
$$ S=e\int d^3 x\;:(\partial_i\phi^\dagger A_i^T\phi-\phi^\dagger A_i^T
\partial_i\phi):\eqno(41)$$
to the lowest order.

	We are now equipped to calculate the scattering amplitudes and cross
sections for several two-body scattering processes. Assuming that the two
particles have initial three-momenta $k$, $l$ and final ones $p$, $q$. The
amplitude will in general has the form
$$A_{fi}=(2\pi)^3 V^{-2}\delta(p+q-k-l) R(p,q,k,l) \eqno(42)$$
where the function $R(p,q,k,l)$ depends on the particular process. By the
method similar to that used in QED [12], the cross section (in two spatial
dimensions it may be more appropriately called the cross width) can be
shown to be
$$\sigma={1\over 2\pi|{\bf k}/k_0-{\bf l}/l_0|}\int d\theta\;{p_0 |R|^2\over
\partial(p_0+q_0)/\partial p_0}\eqno(43)$$
where $\theta$ is the angle between {\bf p} and {\bf k}. This is subject to
the condition
$$ {\bf p}+{\bf q}={\bf k} +{\bf l}\eqno(44a)$$
by which $q_0$ depends on $p_0$ (or $|{\bf p}|$) and $\theta$. In evaluating
the partial derivative $\partial q_0/\partial p_0$, $\theta$ is treated
as a constant. Finally the result (43) is still subject to the additional
condition
$$ p_0+q_0=k_0+l_0\eqno(44b)$$
which, together with (44a), exhibit the conservation of the total momentum
and energy.

	For particle-particle scattering, we have
$$ |i\rangle=a_{\bf k}^\dagger a_{\bf l}^\dagger|0\rangle,\quad
|f\rangle=a_{\bf p}^\dagger a_{\bf q}^\dagger|0\rangle.\eqno(45)$$
The amplitude is given by (42) with
$$ R(p,q,k,l)={e^2\over\kappa}\left[{\epsilon^{ij}k^ip^j(l_0+q_0)\over
2\sqrt{p_0q_0k_0l_0}({\bf p}-{\bf k})^2}+(k\leftrightarrow l)+
(p\leftrightarrow q)+(k\leftrightarrow l,p\leftrightarrow q)\right].
\eqno(46)$$
In achieving this result we have used
$$ \int d{\bf x}\;\epsilon^{ij}\partial_j G({\bf x})e^{i{\bf k\cdot x}}=
i{\epsilon^{ij}k^j\over {\bf k}^2}\eqno(47)$$
which can be verified straightforwardly. In the center-of-mass system, we
attach a subscript $c$ to any quantity and have
$${\bf l}_c=-{\bf k}_c, {\bf q}_c=-{\bf p}_c,\eqno(48a)$$
$$ q_{c0}=p_{c0}=l_{c0}=k_{c0}.\eqno(48b)$$
It is easy to show that
$$\sigma_c={e^4\over \kappa^2}{1\over 2\pi|{\bf k}_c|}\int_0^\pi
d\theta_c\;\cot^2\theta_c\eqno(49)$$
where the upper bound of integration is $\pi$ instead of $2\pi$ because
the two particles are identical. Note that $d\sigma_c/d\theta_c$
is singular at $\theta_c=0$ and $\theta_c=\pi$. The total cross section
is divergent, which implies that the interaction is a long-range one.
In the laboratory system, we may choose ${\bf l}=0$, and thus $l_0=m$.
Eq.(44a) then reduces to
$$ {\bf p}+{\bf q}={\bf k}\eqno(50a)$$
from which one finds
$$ {\partial q_0\over\partial p_0}={p_0\over q_0}\left(1-{{\bf k\cdot p}
\over {\bf p}^2}\right).\eqno(51)$$
Using Eq.(44b) which for the present case reduces to
$$ p_0+q_0=k_0+m,\eqno(50b)$$
together with (50a) and the on-shell relations we obtain
$$ {\bf k\cdot p}=(k_0+m)(p_0-m).\eqno(52)$$
Substituting (52) into (51) we have
$$ {\partial q_0\over\partial p_0}={p_0(p_0-k_0)\over q_0(p_0+m)}.
\eqno(53)$$
Using (50b) once again we arrive at
$$ {\partial (p_0+q_0)\over\partial p_0}={m(k_0+m)\over q_0(p_0+m)}.
\eqno(54)$$
We also find that
$$ R={e^2\over\kappa}{\epsilon^{ij}k^ip^j\over 2\sqrt{p_0q_0k_0m}}
{p_0-q_0\over(p_0-m)(k_0-p_0)}.\eqno(55)$$
Substituting (54) and (55) into (43) we get
$$\sigma={e^4\over\kappa^2}{1\over 8\pi m^2}{|{\bf k}|\over k_0+m}
\int_0^{\pi/2}d\theta\;{(p_0+m)^2(2p_0-k_0-m)^2\sin^2\theta\over
(p_0-m)(p_0-k_0)^2}.\eqno(56)$$
It should be remarked that the upper bound of integration in (56) is $\pi/2$.
This can be easily justified by taking notice of (52) which implies
$\cos\theta\ge0$ and hence $-\pi/2\le\theta\le \pi/2$, and
remembering the fact that the two particles are identical. The
result (56) is not the final one. Given a $\theta$, {\bf p} and {\bf q}
are completely fixed by (50) and the on-shell relations. Therefore one
should manage to solve $p_0$ in $k_0$ and $\theta$. From (52) it is not
difficult to find that
$$p_0={m[m\sin^2\theta+k_0(1+\cos^2\theta)]\over k_0\sin^2\theta+
m(1+\cos^2\theta)}.\eqno(57)$$
Substituting (57) into (56) we arrive at the final result
$$\sigma={e^4\over\kappa^2}{1\over 4\pi m |{\bf k}|}\int_0^{\pi/2}d\theta\;
{[(k_0+m)\sin^2\theta-2m\cos^2\theta]^2\over \sin^2\theta\cos^2\theta
[(k_0+m)\sin^2\theta+2m\cos^2\theta]}.\eqno(58)$$
The singular points of $d\sigma/d\theta$ are $\theta=0$ and $\theta=
\pi/2$, corresponding to  $\theta_c=0$ and $\theta_c=
\pi$, respectively (see Eq.(63)).

	Although explicit Lorentz invariance is lost in the Hamiltonian
formalism, the theory is Lorentz invariant in its original form. Therefore
one naturally expects the result
$$ \sigma=\sigma_c.\eqno(59)$$
To verify this, one must find the relation between $\theta$ and $\theta_c$.
By means of the relations $k\cdot l=k_c\cdot l_c$, $p\cdot l=p_c\cdot l_c$,
and $p\cdot k=p_c\cdot k_c$ and (48) one finds
$$ |{\bf k}_c|=\sqrt{m(k_0-m)\over 2},\eqno(60)$$
$$ |{\bf p}|\cos\theta={|{\bf k}|\over2}(1+\cos\theta_c).\eqno(61)$$
Noting that ${\bf p}_\perp$ (the component of {\bf p} in the direction
perpendicular to {\bf k}) has the same value in the two systems one has
$$ |{\bf p}|\sin\theta=|{\bf k}_c|\sin\theta_c.\eqno(62)$$
On account of these result we obtain
$$\tan\theta=\sqrt{2m\over k_0+m}\tan{\theta_c\over2}.\eqno(63)$$
This is the relation between $\theta$ and $\theta_c$. When $|{\bf k}|\ll m$
it reduces to the nonrelativistic result $\theta_c=2\theta$.
Eqs.(63) and (60)
are sufficient to establish Eq.(59).

	For antiparticle-antiparticle scattering, the result is completely
the same as that of particle-particle scattering. Finally for particle-%
antiparticle scattering, we have
$$ |i\rangle=a_{\bf k}^\dagger b_{\bf l}^\dagger|0\rangle,\quad
|f\rangle=a_{\bf p}^\dagger b_{\bf q}^\dagger|0\rangle.\eqno(64)$$
The amplitude is given by (42) with
$$R(p,q,k,l)={e^2\over \kappa}{1\over 2\sqrt{p_0q_0k_0l_0}}\times$$
$$\times
\left[{(q_0+l_0)\epsilon^{ij}p^ik^j+(p_0+k_0)\epsilon^{ij}q^il^j\over
({\bf p}-{\bf k})^2}+{(k_0-l_0)\epsilon^{ij}p^iq^j+(q_0-p_0)\epsilon^{ij}
k^il^j\over ({\bf k}+{\bf l})^2}\right].\eqno(65)$$
In the	center-of-mass system, the second term in the square bracket is not
well-defined. So we begin with ${\bf l}_c=-{\bf k}_c+\varepsilon {\bf k}_c$
and finally let $\varepsilon\to 0$. The result can be easily shown to be
$$\sigma_c={e^4\over\kappa^2}{1\over 8\pi|{\bf k}_c|}\int_{-\pi}^\pi
d\theta_c\;\left[1+{{\bf k}_c^2\over k_{c0}^2}\sin^2{\theta_c\over2}
\right]^2\cot^2{\theta_c\over2}.\eqno(66)$$
In the laboratory system, we choose ${\bf l}=0$ and find
$$ R={e^2\over\kappa}{\epsilon^{ij}p^ik^j\over 2\sqrt{p_0q_0k_0m}}
{q_0+k_0\over (q_0-m)(k_0+m)}.\eqno(67)$$
Substituting (67) and (54) into (43) yields
$$\sigma={e^4\over\kappa^2}{1\over 8\pi m^2}{(k_0-m)^{1/2}\over(k_0+m)^{5/2}}
\int_{-\pi/2}^{\pi/2}d\theta\;{(p_0+m)^2(p_0-m)(2k_0-p_0+m)^2\sin^2\theta
\over (p_0-k_0)^2}.\eqno(68)$$
The range of $\theta$ has been explained before. In the present case the
two particles are distinguishable, so the integration is performed over
the full range of $\theta$. In (68) $p_0$ depends on $k_0$ and $\theta$,
which can be eliminated by employing (57). The final result turns out to be
$$\sigma={e^4\over\kappa^2}{4m\over\pi|{\bf k}|}\int_{-\pi/2}^{\pi/2}
d\theta\;{(k_0\sin^2\theta+m\cos^2\theta)^2\cot^2\theta\over
[k_0\sin^2\theta+m(1+\cos^2\theta)]^3}.\eqno(69)$$
It is easily seen that $d\sigma/d\theta$ is  singular at $\theta=0$.
At $\theta=\pm\pi/2$, however, it vanishes.
This means that the incident particle
can not transfer all its kinetic energy to the antiparticle, and might
imply that there exists some attractive force between the particle and the
antiparticle. The total cross section is again divergent. Two-body
potentials among particles and antiparticles are under investigation,
which will exhibit the property of the interaction force more explicitly.
Using Eqs.(60) and (63), we once again achieve the result (59). If there
exists some system of scalar particles with pure Chern-Simons interaction,
one may examine the results (58) and (69) by experiments. Extension of the
present work to the Chern-Simons theory interacting with a spinor field
is currently under progress. Higher-order corrections to the cross sections
are also being studied. Results will be reported subsequently.

\vskip 2pc

	The author thanks Professor Qi-zhou Chen and Professor
Guang-jiong Ni for their encouragement. This work was supported by
the Doctoral Foundation of the National Education Commission of China.

\end{document}